\documentclass[12pt]{article}
\usepackage[dvips]{graphicx}
\usepackage{epsfig}
\usepackage{amsmath,amssymb,amsbsy}
\usepackage{amsfonts}

\begin{document}
\thispagestyle{empty}
\begin{center}

{\Large\bf{
Properties of the light quarks and antiquarks\\ 
\vskip 0.4cm
in the statistical approach}}
\vskip1.4cm
{\bf Claude Bourrely}
\vskip 0.3cm
Aix Marseille Univ, Univ Toulon, CNRS, CPT, Marseille, France\\
\vskip 0.5cm
{\bf Jacques Soffer}
\vskip 0.3cm
Physics Department, Temple University,\\
1925 N, 12th Street, Philadelphia, PA 19122-1801, USA
\vskip 0.5cm
{\bf Abstract}
\end{center}

In the quantum statistical parton distributions approach proposed more than one
decade ago, one imposes relations between quarks and antiquarks expressions, 
which lead
to very specific properties for the antiquarks. These properties have been verified
up to now by recent data and it is a real challenge also for forthcoming experimental results,
mainly in the high $x$ region.

\clearpage
\newpage

Let us now recall the main features of the statistical approach for building
up the parton distributions function (PDFs). In this approach  
we treat simultaneously unpolarized distributions and helicity distributions, 
a unique situation in the literature.

The fermion distributions are given by the sum of two terms,
a quasi Fermi-Dirac function and a helicity independent diffractive
contribution:
\begin{equation}
xq^h(x,Q^2_0)=
\frac{A_{q}X^h_{0q}x^{b_q}}{\exp [(x-X^h_{0q})/\bar{x}]+1}+
\frac{\tilde{A}_{q}x^{\tilde{b}_{q}}}{\exp(x/\bar{x})+1}~,
\label{eq1}
\end{equation}
\begin{equation}
x\bar{q}^h(x,Q^2_0)=
\frac{{\bar A_{q}}(X^{-h}_{0q})^{-1}x^{\bar{b}_ q}}{\exp
[(x+X^{-h}_{0q})/\bar{x}]+1}+
\frac{\tilde{A}_{q}x^{\tilde{b}_{q}}}{\exp(x/\bar{x})+1}~,
\label{eq2}
\end{equation}
at the input energy scale $Q_0^2=1 \mbox{GeV}^2$. 
For the antiquarks we propose the ansatz (\ref{eq2}) perfectly compatible
with experiemental data.
We note that the diffractive
term is absent in the quark helicity distribution $\Delta q$, in the quark
valence contribution $q - \bar q$ and in $u - d$.\\
In Eqs.~(\ref{eq1},\ref{eq2}) the multiplicative factors $X^{h}_{0q}$ and
$(X^{-h}_{0q})^{-1}$ in
the numerators of the non-diffractive parts of the $q$'s and $\bar{q}$'s
distributions, imply a modification
of the quantum statistical form, we were led to propose in order to agree with
experimental data.
The parameter $\bar{x}$ plays the role of a {\it universal temperature}
and $X^{\pm}_{0q}$ are the two {\it thermodynamical potentials} of the quark
$q$, with helicity $h=\pm$. They represent the fundamental parameters of
the model. Notice the change of sign of the potentials
and helicity for the antiquarks. For a given flavor $q$ the corresponding quark and antiquark distributions
involve the free parameters, $X^{\pm}_{0q}$, $A_q$, $\bar {A}_q$,
$\tilde {A}_q$, $b_q$, $\bar {b}_q$ and $\tilde {b}_q$, whose number is reduced to $\it
seven$ by the valence sum rule, $\int (q(x) - \bar
{q}(x))dx = N_q$, where $N_q = 2, 1 ~~\mbox{for}~~ u, d$, respectively.

From a fit of unpolarized and polarized experimental data we have obtained
for the potentials the values \cite{Bourrely:2015kla}:
\begin{equation}
X_u^+ = 0.475, \quad X_u^- = X _d^- = 0.307,  \quad X_d^+ = 0.244. 
\label{potval}
\end{equation}
To our surprise,
it turns out that two potentials have identical numerical values,
so  for light quarks we have found  the following hierarchy between the 
different potential components 
\begin{equation}
X_u^+ >  X_u^- = X _d^- >  X_d^+ .
\label{potherar}
\end{equation}
We notice that quark helicity PDFs increases with the potential value, while antiquarks
helicity PDFs increases when the potential decreases.

As a consequence of the above hierarchy it follows an hierarchy on the quarks 
helicity distributions,
\begin{equation}
 xu_+(x) > xu_-(x) = xd_-(x) > xd_+(x)
\label{ineq}
\end{equation}
and an obvious hierarchy for the antiquarks, namely
\begin{equation}
x\bar d_- (x) >  x\bar d_+ (x) = x\bar u_+ (x) >  x\bar u_- (x) ,
\label{ineqbar}
\end{equation}

\begin{figure}[hbp]   
\vspace*{-20.5ex}
\begin{center}
\includegraphics[width=8.0cm]{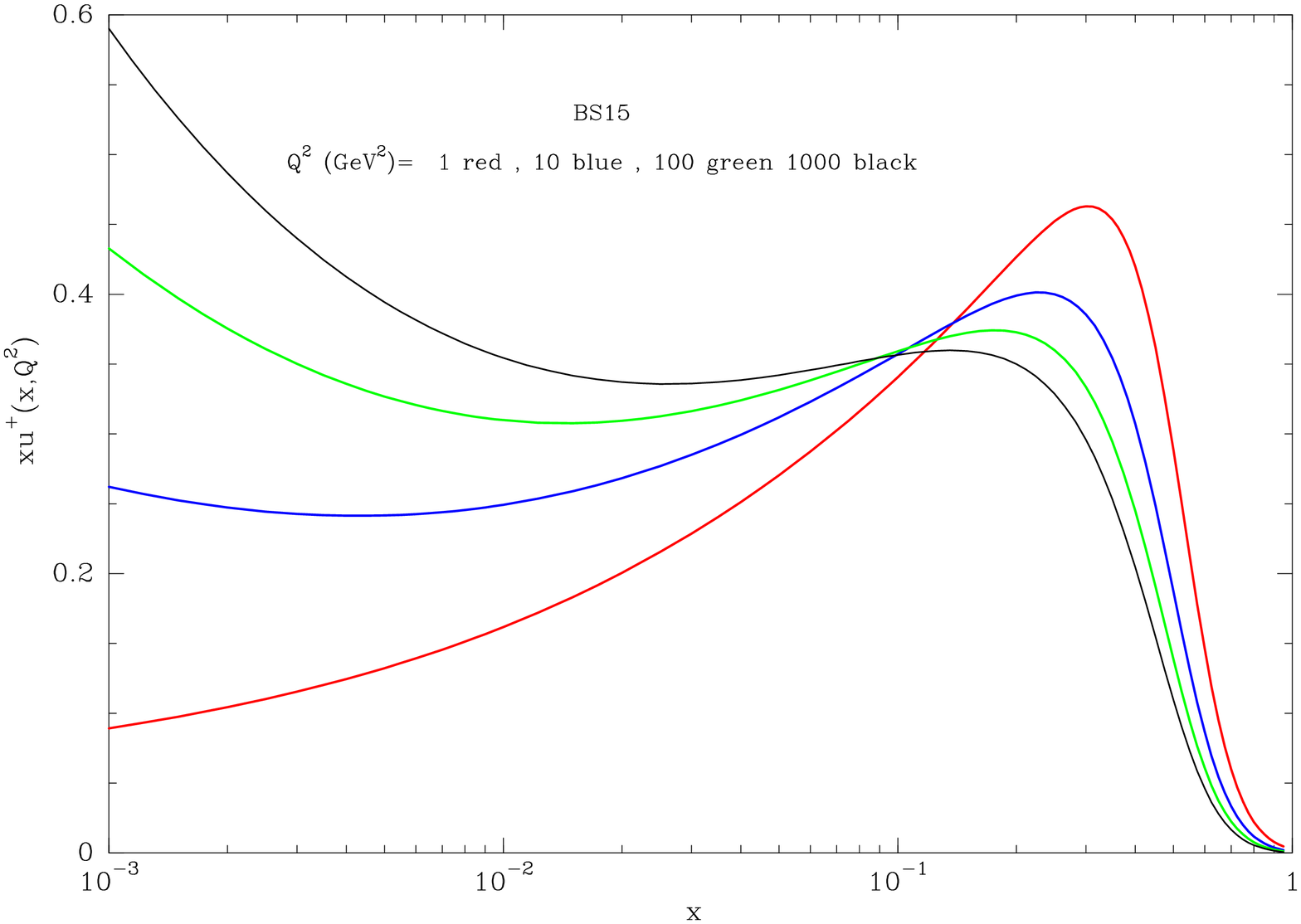}
\vspace*{+3.0ex}
\includegraphics[width=8.0cm]{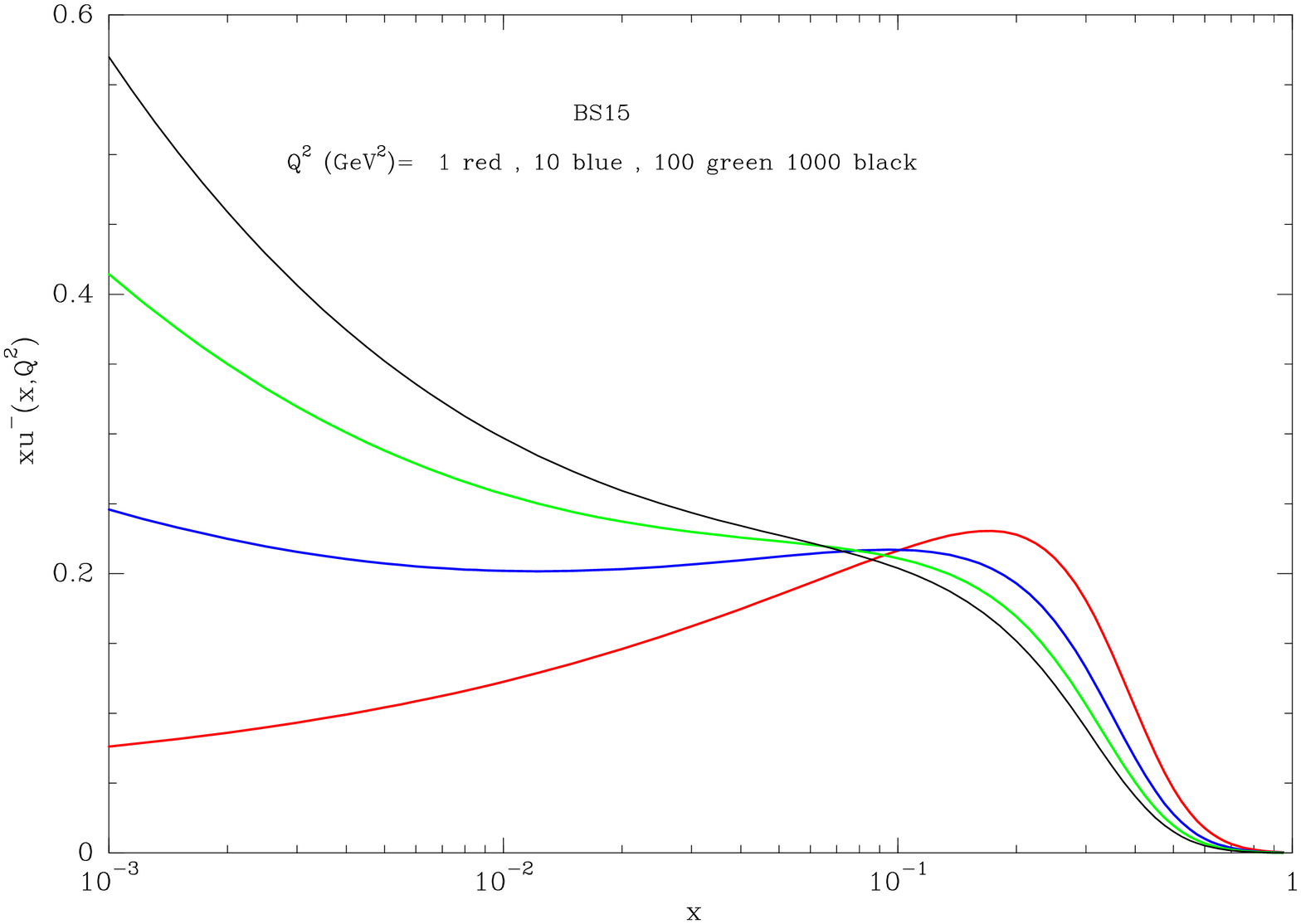}
\includegraphics[width=8.0cm]{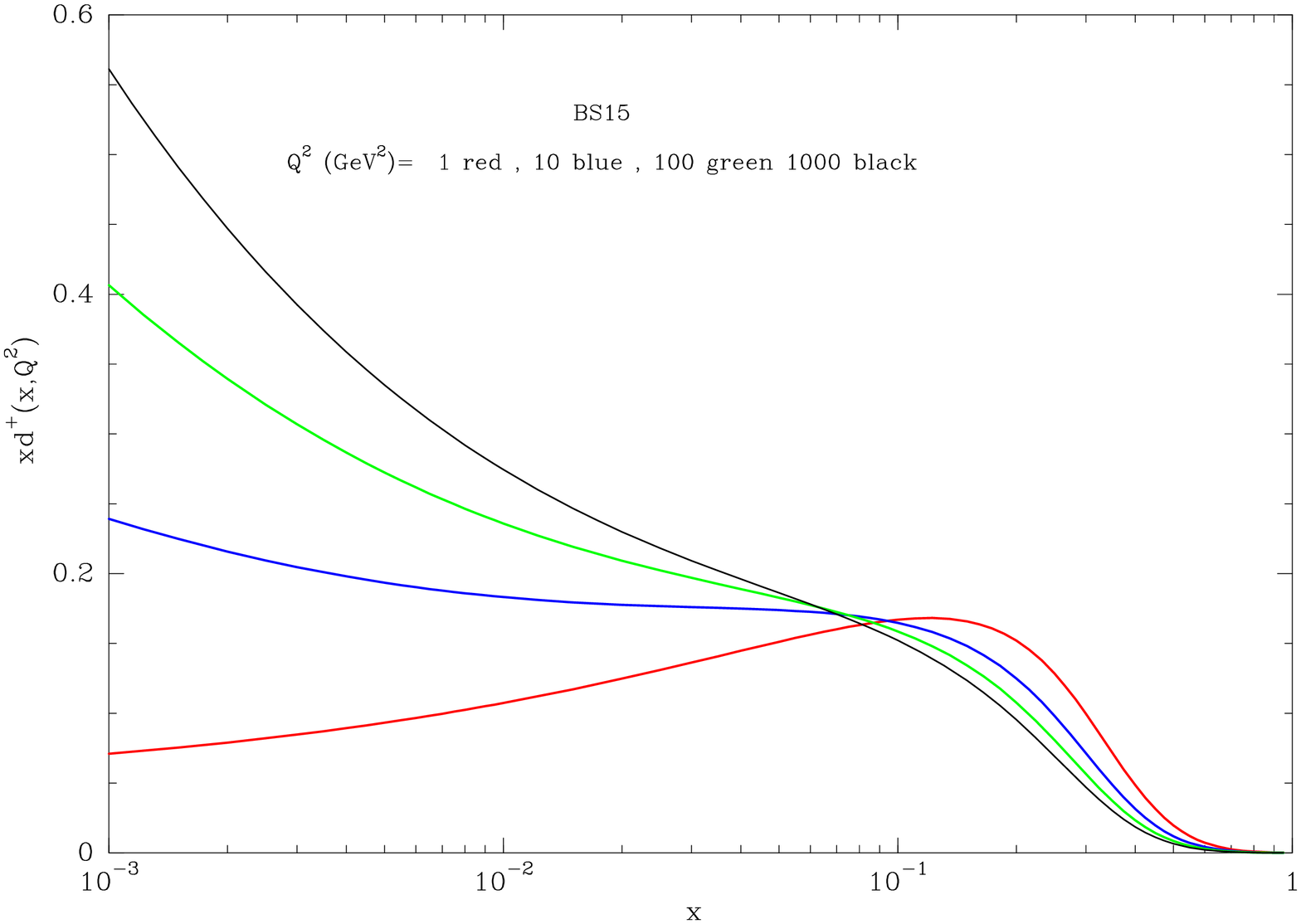}
\vspace*{+1.0ex}
\caption[*]{\baselineskip 1pt
 The different helicity components of the light quark distributions $xf(x,Q^2)$ ($f=u_+, u_- = d_-, d_+, \mbox {from top to bottom}$), versus $x$, at $Q^2=10, 100, 1000 \mbox{GeV}^2$,
after NLO QCD evolution, from the initial scale $Q^2 = 1 \mbox{GeV}^2$.}
\label{quarkhel}
\end{center}
\end{figure}

\begin{figure}[hbp]   
\vspace*{-20.5ex}
\begin{center}
\includegraphics[width=8.0cm]{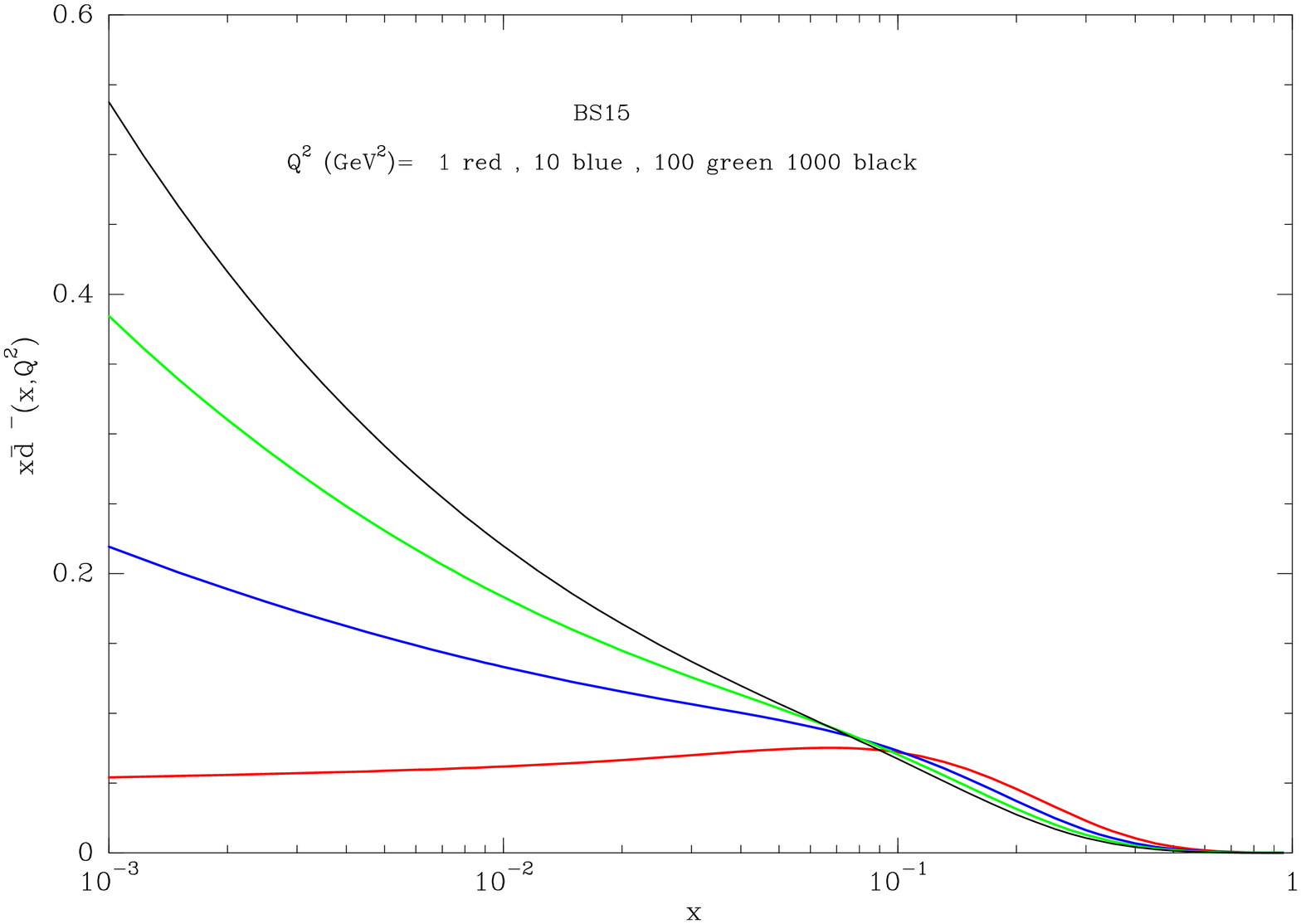}
\vspace*{+3.0ex}
\includegraphics[width=8.0cm]{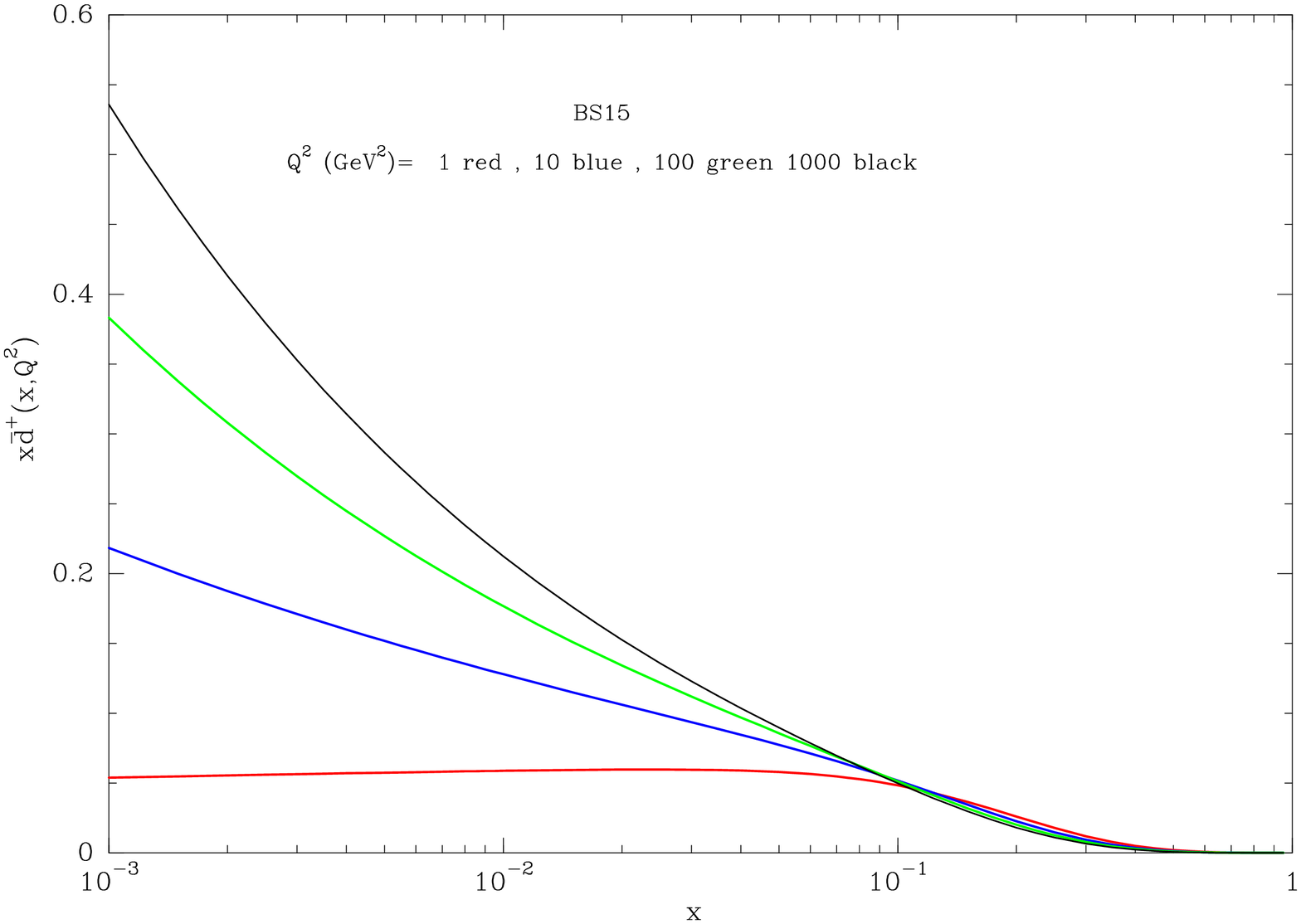}
\vspace*{-2.0ex}
\includegraphics[width=8.0cm]{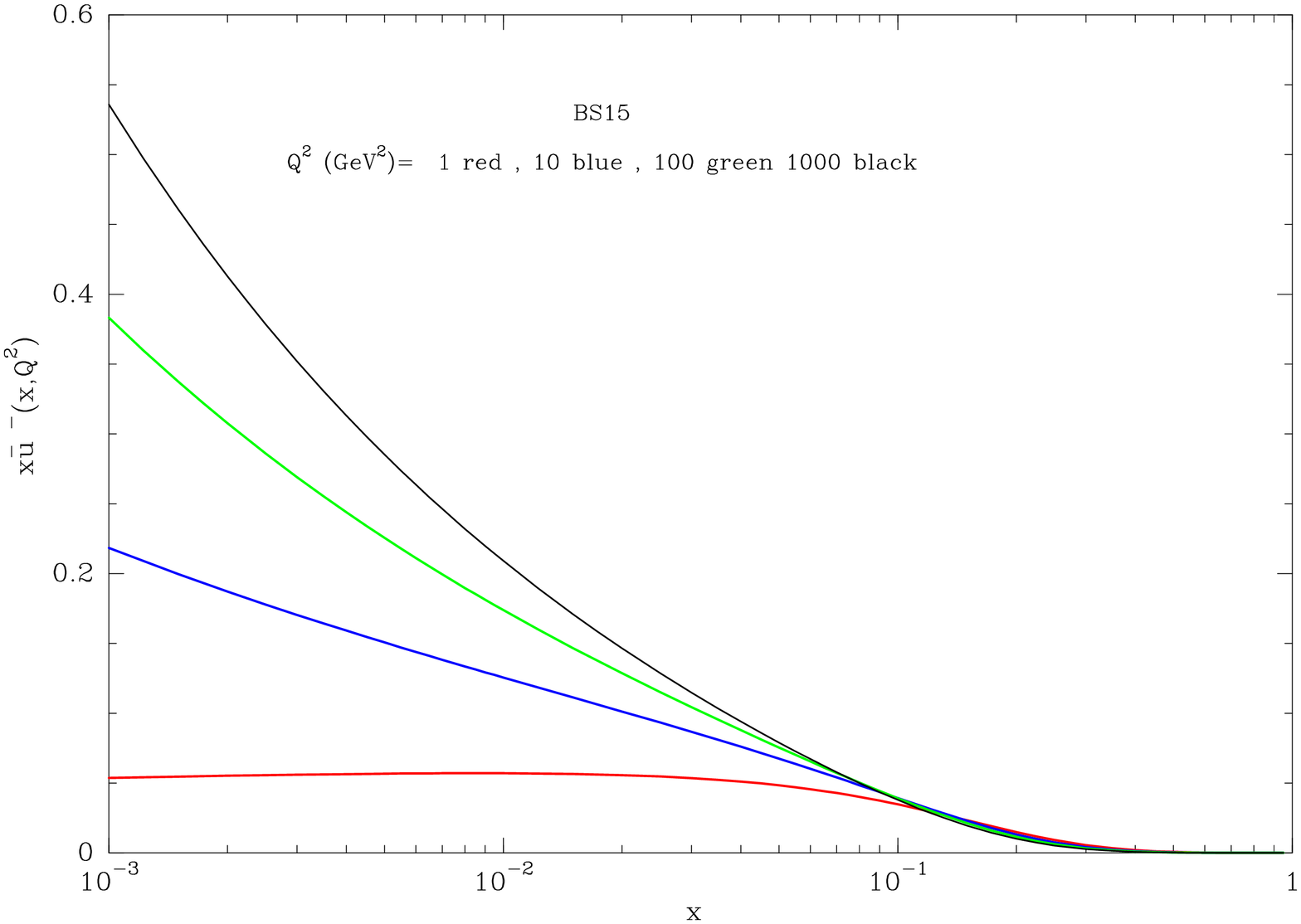}
\vspace*{+1.0ex}
\caption[*]{\baselineskip 1pt
The different helicity components of the light antiquark distributions $xf(x,Q^2)$ ($f=\bar d_-, \bar d_+ = \bar u_+, \bar u_-, \mbox {from top to bottom}$), versus $x$, at $Q^2=10, 100, 1000 \mbox{GeV}^2$,
after NLO QCD evolution, from the initial scale $Q^2 = 1 \mbox{GeV}^2$.}
\label{antiquarkhel}
\end{center}
\end{figure}

It is important to note that these inegalities Eqs. (\ref{ineq})-(\ref{ineqbar})
 are 
preserved by the next-to-leding QCD evolution, 
as we can see on Fig. \ref{quarkhel} and Fig. \ref{antiquarkhel}, at least 
outside the diffractive region.\\
One important remark is that we have checked that the initial analytic form  
Eqs.~(\ref{eq1},\ref{eq2}), is almost preserved by the $Q^2$ evolution with some small changes of the parameters.
\newpage

\clearpage
\newpage

\begin{figure}[htp]   
\vspace*{-20.5ex}
\begin{center}
\includegraphics[width=6.5cm]{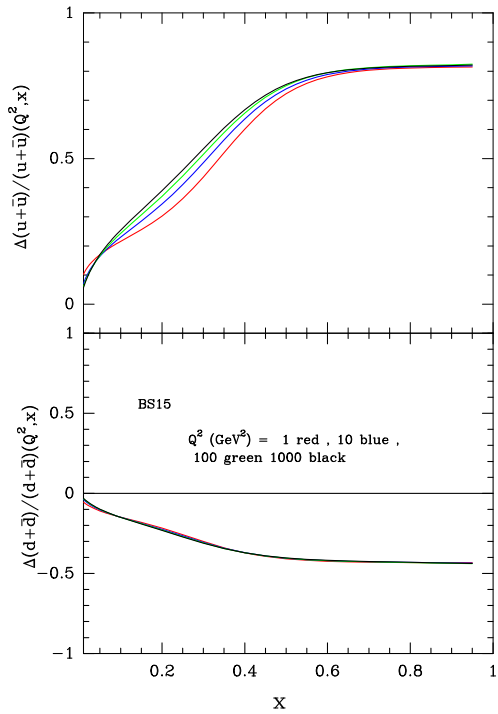}
\caption[*]{\baselineskip 1pt
Predicted ratios $(\Delta u(x,Q^2) + \Delta \bar u(x,Q^2)) /  (u(x,Q^2 + \bar u(x,Q^2))$ and
$(\Delta d(x,Q^2) + \Delta \bar d(x,Q^2)) /  (d(x,Q^2 + \bar d(x,Q^2))$, versus $x$, at $Q^2=10, 100, 1000 \mbox{GeV}^2$,}
\label{disratiosl}
\end{center}
\end{figure}

One clearly concludes that $u(x,Q^2) > d(x,Q^2)$ implies a flavor-asymmetric light sea, i.e. $\bar d(x,Q^2) > \bar u(x,Q^2)$, a trivial consequence of the Pauli 
exclusion principle, which is built in. Indeed this is based on the fact that the proton contains two $u$ quarks and only one $d$ quark.\\ 
Let us move on to mention more significant consequences concerning the helicity
 distributions which follow from Eqs. (\ref{potval})-(\ref{ineqbar}).
First for the $u$-quark 
\begin{equation}
 x\Delta u(x,Q^2) > 0 \quad\quad  x\Delta \bar u(x,Q^2) > 0.
\end{equation}
Similarly for the $d$-quark 
\begin{equation}
 x\Delta d(x,Q^2) < 0 \quad\quad  x\Delta \bar d(x,Q^2) < 0.
\end{equation}
We have made these predictions almost 15 years ago \cite{bbs1} when, for simplifying reasons, it was more natural to assume that   $x\Delta \bar u(x,Q^2) = x\Delta \bar d(x,Q^2)$.\\
Our predicted signs and magnitudes have been confirmed  \cite{Bourrely:2015kla} by the measured single-helicity asymmetry $A_L$ in the $W^{\pm}$ production at BNL-RHIC from STAR \cite{Adamczyk:2014xyw}.\\
Another important earlier prediction concerns the Deep Inelastic Scattering (DIS) asymmetries, more precisely $(\Delta u(x,Q^2) + \Delta \bar u(x,Q^2)) /  (u(x,Q^2 + \bar u(x,Q^2))$ and
$(\Delta d(x,Q^2) + \Delta \bar d(x,Q^2)) /  (d(x,Q^2) + \bar d(x,Q^2))$, shown in Fig. \ref{disratiosl}. Note that the  data, so far, are in agreement with these predictions. In the high $x$ region they differ from those which impose, for both quantities, the value one for $x=1$. This is another challenge, since only up to $x=0.6$, they have  been measured at JLab
\cite{JLab}. 

\begin{figure}[htp]   
\vspace*{-21.50ex}
\begin{center}
\includegraphics[width=8.0cm]{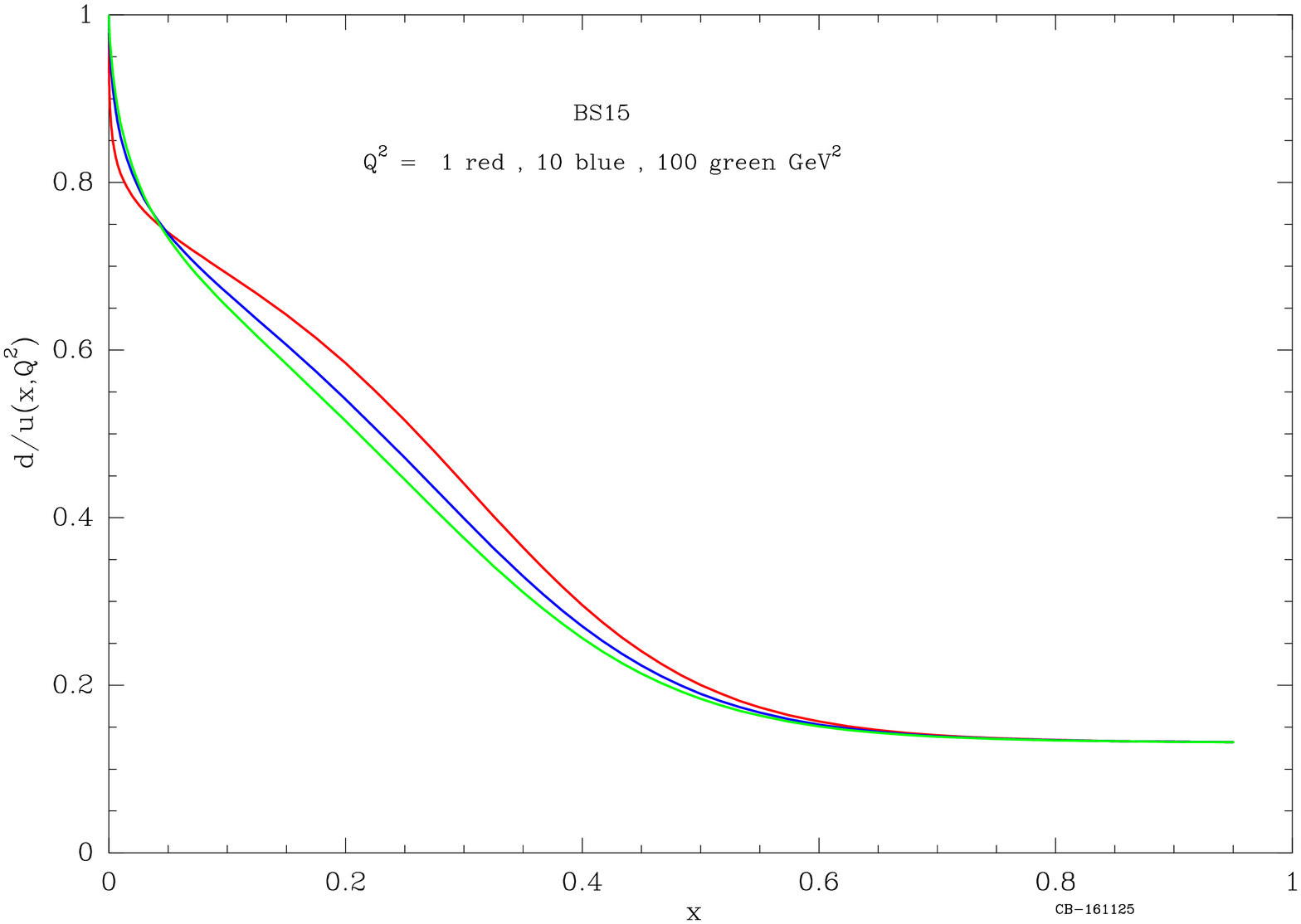}
\vspace*{+7.0ex}
\includegraphics[width=8.0cm]{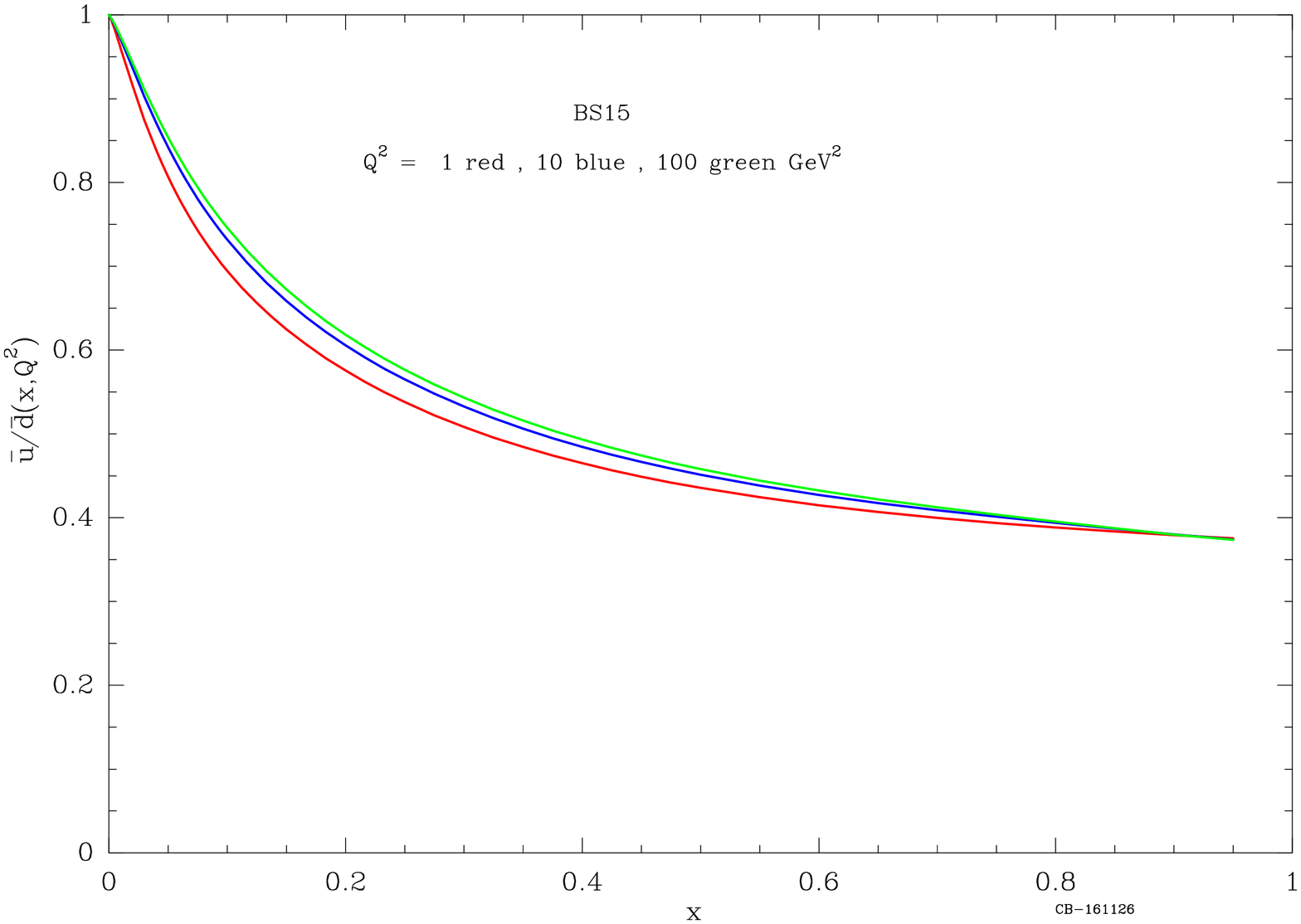}
\vspace*{-5.0ex}
\caption[*]{\baselineskip 1pt
The ratios $d(x,Q^2)/u(x,Q^2)$ (${\it top}$) and $\bar u(x,Q^2)/\bar d(x,Q^2)$ (${\it bottom}$) versus $x$ for different values of $Q^2$.}
\label{ratios}
\end{center}
\end{figure}

There are two more important consequences which relate unpolarized and helicity distributions, namely for quarks
\begin{equation}
xu(x,Q^2) -  xd(x,Q^2) = x\Delta u(x,Q^2) -  x\Delta d(x,Q^2)   > 0 ,
\label{relatqdq}
\end{equation}
and similarly for antiquarks
\begin{equation}
x\bar d(x,Q^2) - x\bar u(x,Q^2) = x\Delta \bar u(x,Q^2) -  x\Delta \bar d(x,Q^2) > 0
 .
\label{relatqdqbar}
\end{equation}
This means that the flavor asymmetry of the light antiquark distributions is the same for the corresponding helicity distributions, as noticed long time ago \cite{bbs-rev}.\\
Now let us come back to all these components $xu_+ (x, Q^2), ...x\bar u_- (x,Q^2)$ and more precisely to their $x$-behavior. It is clear that $xu_+(x,Q^2)$ is the largest one and they are all 
monotonic decreasing functions of $x$ at least for $x > 0.2$, outside the region dominated by the diffractive contribution see Figs. \ref{quarkhel}-\ref{antiquarkhel}.\\
 Similarly $x\bar d_- (x\,Q^2)$ is the largest of the antiquark components.\\
Therefore if one considers the ratio $xd(x,Q^2)/xu(x,Q^2)$, its value is one at $x=0$, because the diffractive contribution dominates and, due to the monotonic decreasing, it decreases for increasing $x$.\\
This falling $x$-behavior has been verified experimentaly from the ratio of the DIS structure functions $F_{2}^{d} / F_{2}^p$ and the charge asymmetry of the $W^{\pm}$ production in $\bar p p$ collisions \cite{Kuhlmann:1999sf}.\\
Similarly if one considers the ratio  $x\bar u(x,Q^2) /x\bar d(x,Q^2)$, its value is one at $x=0$, because the diffractive contribution dominates and, due to the monotonic decreasing, it decreases for increasing $x$.\\
By looking at the curves ( See Figure \ref{ratios}), one sees similar behaviors. In both cases in the vicinity of $x=0$ one has a sharp behavior due to the fact that the diffractive contribution dominates and in the high $x$ region
there is a flattening out above $x \simeq 0.6$. It is remarkable to see that these ratios have almost no $Q^2$ dependence.\\
To conclude we predict a monotonic increase of the ratio $x\bar d (x,Q^2) /x\bar u(x,Q^2)$. This was first observed in the low $x$ region by the E866/NuSea collaboration \cite{E866} and very recently there is a serious indication
from the preliminary results of the SeaQuest collaboration, that this trend persists beyond $x=0.2$ \cite{reimer}. \\

This prediction results from the following characteristic features of the statistical approach:\\
1) The hierarchy of the potentials Eq. (3) which are fundamental parameters in the approach.\\
2) The monotonic decreasing with $x$ which is related to the Fermi-Dirac 
expression used to parameterise the parton distributions.\\
3) The expressions between quark and antiquarks 
we have supposed and which allow to relate the behavior of the ratios  
$xd(x,Q^2)/xu(x,Q^2)$ and $x\bar u(x,Q^2) /x\bar d(x,Q^2)$.\\
Due to the high predictive power of our model  it is a real challenge
for several forthcoming data.

\clearpage
\newpage


\begin{thebibliography}{99}

\bibitem{Bourrely:2015kla}
C.~Bourrely, J.~Soffer, Nucl. Phys. A {\bf 941} 307 (2015).

\bibitem{bbs1} 
C. Bourrely, F. Buccella and J. Soffer, Eur. Phys. J. C {\bf 23}, 487 (2002).

\bibitem{Adamczyk:2014xyw}
L. Adamczyk, et al., STAR Collaboration,  Phys. Rev. Lett. {\bf 113}, 072301 (2014).

\bibitem{JLab}
X, Zheng et al.,  Phys. Rev. C {\bf 70}, 065207 (2004).
      

\bibitem{bbs-rev} 
C.~Bourrely, J.~Soffer and F.~Buccella, Eur.\ Phys.\ J.\ C {\bf 41}, 327 (2005).

\bibitem{Kuhlmann:1999sf}
S. Kuhlmann, et al., Phys. Lett. B {\bf 476}, 291 (2000).

\bibitem{E866} FNAL Nusea Collaboration, E. A. Hawker {\it et al.},
Phys. Rev. Lett. {\bf 80}, 3715 (1998); J. C. Peng {\it et al.},
Phys. Rev. D {\bf 58}, 092004 (1998).

\bibitem{reimer}
P. Reimer, Invited talk at "DIFFRACTION 2016", Sept. 02 - 08, 2016, Acireale, Sicily (Italy), AIP Conference Proceedings (2017).



\end{thebibliography}
\end{document}